# Stark effect spectroscopy of mono- and few-layer MoS$_2$

**AUTHOR NAMES**.

*J. Klein[1*], J. Wierzbowski[1*], A. Regler[1,2], J. Becker[1], F. Heimbach[3], K. Müller[1,4], M. Kaniber[1†] and J. J. Finley[1†]*

**AUTHOR ADDRESS**.

[1] Walter Schottky Institut und Physik Department, Technische Universität München, Am Coulombwall 4, 85748 Garching, Germany

[2] Institute for Advanced Study, Technische Universität München, Lichtenbergstrasse 2a, 85748 Garching, Germany

[3] Lehrstuhl für Physik funktionaler Schichtsysteme, Physik Department E10, Technische Universität München, James-Franck-Straße 1, 85747 Garching, Germany

[4] E. L. Ginzton Laboratory, Stanford University, Stanford, CA 94305, USA

* These authors contributed equally

**ABSTRACT**. We demonstrate electrical control of the A-exciton interband transition in mono- and few-layer MoS$_2$ crystals embedded into photocapacitor devices via the DC Stark effect. Electric field dependent low-temperature photoluminescence spectroscopy reveals a significant tuneability of the

A-exciton transition energy, up to ~ 16 meV, from which we extract the mean DC exciton polarisability $\langle \bar{\beta} \rangle_N = (0.58 \pm 0.25) \times 10^{-8}$ DmV$^{-1}$. The exciton polarisability is shown to be layer-independent, indicating a strong localisation of both electron and hole wave functions in each individual layer.

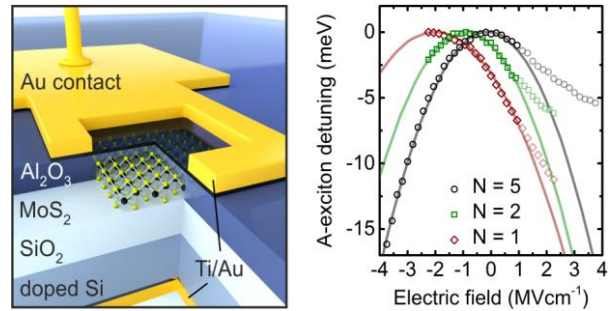

**KEYWORDS**. 2D materials, transition metal dichalcogenides, molybdenum disulfide, quantum confined Stark effect, electrical control, exciton polarisability

Atomically thin two-dimensional materials like graphene[1] pave the way for a wealth of applications ranging from highly efficient optoelectronic devices[2] to energy harvesting[3]. In particular, the semiconducting transition metal dichalcogenides (TMDCs) such as $MoS_2$, $MoSe_2$, $WS_2$ and $WSe_2$ are of particular interest due to their direct electronic band gap in the monolayer limit[4,5] and their strong valley optical selection rules in the absence of inversion symmetry[6,7]. $MoS_2$ is probably the most widely investigated of these materials and it has already been utilised in its few layer form in a variety of prototype electronic and optoelectronic devices. Examples include transistors[8] with on/off ratios ≥ $10^8$ or ultra-sensitive photodetectors[9] with a photoresponsivity exceeding 800 AW$^{-1}$. For such devices, the ability to control the optical response using a convenient control parameter provides much needed flexibility for switching interactions and spectrally tuning of the absorption and emission edge. Tuneability of the emission energy of the A-exciton in $MoS_2$ has been experimentally demonstrated by introducing strain[10], by doping few-layer $MoS_2$ crystals using proximal photoswitchable molecules[11] or using lateral electric fields in field effect devices[12]. Moreover, spectral tuning of the energy gap using vertically applied electric fields has been theoretically proposed for bilayer $MoS_2$ crystals.[13-16]

In this letter, we report on electrical control of the A-exciton emission energy of mono- and few-layer $MoS_2$ crystals using the DC Stark effect. Hereby, we employ a lithographically defined micro-capacitor device which facilitates optical access to atomically thin $MoS_2$ crystals whilst the electric field perpendicular to the basal plane of the crystal is tuned. This device allows to perform electric field controlled photoluminescence (PL) spectroscopy and to extract the hitherto unexplored A-exciton polarisabilities. Moreover, we observe a nonzero excitonic dipole moment that increases as the layer number reduces.

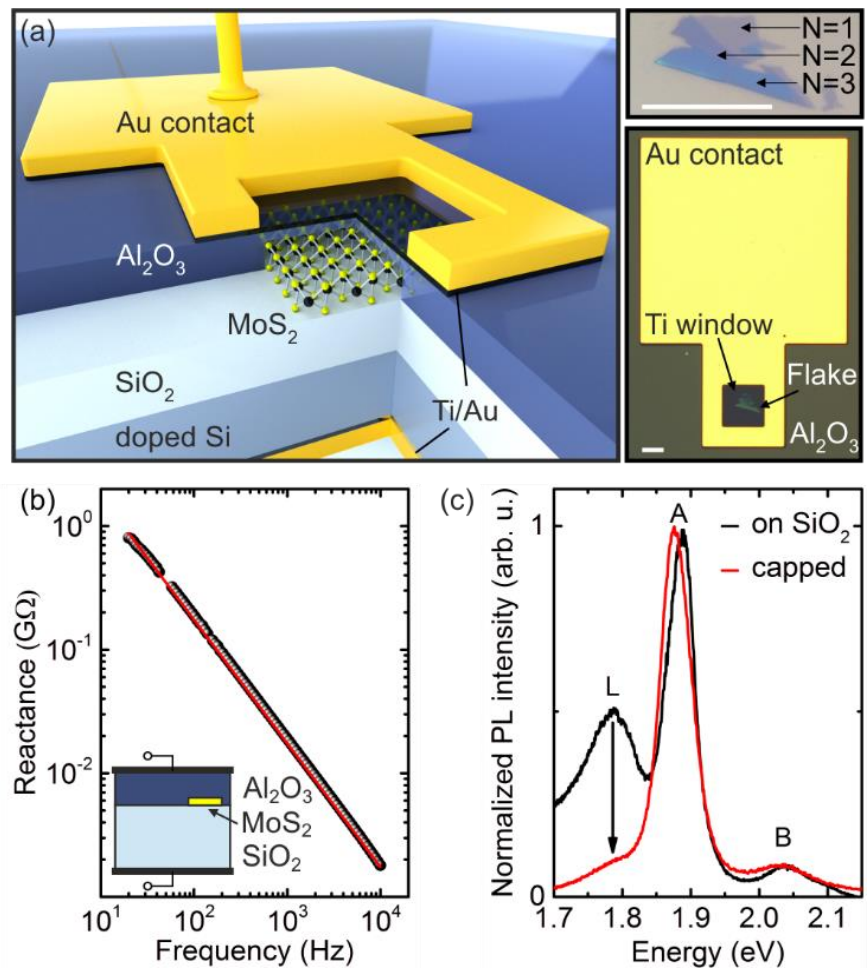

**Figure 1**. (a) Schematic illustration of the $MoS_2$ micro-capacitor devices investigated in this letter. The cross-sectional cut illustrates the layer sequence. Right panels: Optical microscope images of the as-prepared N = 1, 2, 3 crystals before device fabrication and an optical microscope image of an as-fabricated device top contact. The scale bars are 20 μm in both images. (b) Typical frequency dependence measurement of the capacitive reactance (black points) of the simplified capacitor model (inset). The red curve represents a $1/(2\pi fC)$ fit to the data. (c) Normalised low-temperature (10 K)

photoluminescence spectrum recorded from the N = 1 MoS$_2$ region on SiO$_2$ (black curve) and the same region capped with Al$_2$O$_3$ on (red curve) as discussed in the main text. Both spectra reveal A- and B-exciton transitions and the low energy L peak.

The micro-capacitor device structure investigated is schematically illustrated in figure 1(a) and consists of a few-layer MoS$_2$ crystal sandwiched between asymmetric dielectric layers and metal contacts. The device was fabricated by mechanically cleaving commercially available, natural bulk MoS$_2$ (SPI supplies) using the scotch tape method[1] and transferring few-layer MoS$_2$ crystals onto a Si/SiO$_2$ substrate. We used degenerately doped Si substrates acting as global back-gate with 200 nm or 290 nm (> 10$^{19}$ cm$^{-3}$) thermally grown SiO$_2$. Suitable few-layer crystals that were spatially isolated from bulk material were chosen for further device fabrication. The layer number in each MoS$_2$ micro-crystallite was determined using Raman spectroscopy[17] (see Supporting Information section S1) and by assessment of the phase contrast from microscope images[18]. Throughout this letter we solely investigate mono- (N = 1) to penta-layer (N = 5) crystals.

An optical microscope image of a typical as-prepared flake is shown in figure 1(a) featuring N = 1, 2, 3 MoS$_2$ regions. After transfer, we employed atomic layer deposition (ALD) at 100°C to produce a 20 nm thick Al$_2$O$_3$ ($\varepsilon_r$ = 9.3) capping layer to serve as a high-k dielectric on top of the MoS$_2$ crystals. Following this, a two-step optical lithography and metallisation process was used to deposit a 4.5 nm thick semi-transparent Ti layer and a 20 nm/130 nm thick Ti/Au contact layer. A typical optical microscope image of an as-fabricated device with embedded MoS$_2$ crystals is presented in figure 1(a). The Ti/Au layer has a clear opening of 40 × 40 μm² enabling optical access to selected regions of the atomically thin MoS$_2$ crystals. In a final step, we established the global back contact by removing the residual oxide on the backside of the Si substrate using hydrofluoric acid followed by deposition of 30 nm/130 nm Ti/Au. By applying a voltage between the upper Au-contact and the degenerately doped

Si-substrate, a uniform electric field is developed in the MoS$_2$, oriented perpendicular to the basal plane of the crystal.

In order to verify the functionality of our devices, we measured their capacitance and compared with expectations. For a simple parallel plate capacitor consisting of the two conducting plates (Ti/Au and doped Si) and the two dielectrics (Al$_2$O$_3$ and SiO$_2$) with a few-layered MoS$_2$ crystal sandwiched in between the total device capacitance is estimated to be $C_{theo}$ ~ (7.95 ± 0.39) pF (see Supporting Information section S2) with an error originating from uncertainties in the top-gate area and the thicknesses of the dielectrics. The capacitance of the thin MoS$_2$ crystals can be neglected. We then measured the static reactance $X_C$ as a function of the modulation frequency $f$ in a four-point geometry, with typical data shown in figure 1(b) on a double-logarithmic scale. The total reactance exhibits a clear 1/$f$ dependence as expected for a purely capacitive reactance, with a 1/(2$\pi fC$) fit to the data yielding a capacitance of $C_{exp}$ = (9.12 ± 0.01) pF (red curve) in fair qualitative agreement with our expectations. This observation demonstrates that the device operates as a pure capacitor with negligible leakage currents ($j < 10^{-10}$ µA/µm$^2$) over the entire range of applied voltages (electric fields) explored in this paper (see Supporting Information section S3).

To link the gate voltage applied between the Si-substrate and the top Au-contact and the electric field applied to the MoS$_2$, we performed finite-element simulations (see Supporting Information section S3). The combination of Al$_2$O$_3$ and SiO$_2$ as insulating layers enables the application of large electric fields up to |F| = 3.75 MVcm$^{-1}$ for the maximum applied voltages explored in our experiments (± 100 V). The maximum applied electric fields were chosen to be well below the breakdown fields for both the thin Al$_2$O$_3$ ALD films (~ 17 MVcm$^{-1}$)[19] and thermally grown SiO$_2$ (~ 8 MVcm$^{-1}$),[20] respectively.

We now continue to discuss the optical properties of the embedded MoS$_2$ flakes. For all PL measurements discussed throughout this letter, we excited the samples with a frequency-doubled Nd:YAG laser at 532 nm and an excitation power density < 0.5 kWcm$^{-2}$ in order to prevent heating of the sample and photo-doping effects[21,22]. All optical measurements were performed using a He flow cryostat at 10 K. Typical low-temperature PL spectra recorded from uncapped monolayers of MoS$_2$ on SiO$_2$, without an Al$_2$O$_3$ cap, and capped with Al$_2$O$_3$ (but unbiased) are presented in figure 1(c). The uncapped spectrum (black) exhibits recombination from the A- and B-excitons at $E_A \sim 1.88$ eV and $E_B \sim 2.04$ eV,[5,23] respectively, and a fairly strong emission from the defect activated luminescence L-peak at $E_L \sim 1.79$ eV.[23] Consistent with previous works,[24] we observe a considerable suppression of the L peak for all N = 1 flakes when capped by Al$_2$O$_3$. Despite a red shift of the A-exciton by $-(11 \pm 5)$ meV we observed no significant changes in the line shape or significant quenching of the PL signal. Hence, we conclude that the fabrication process used to produce the photocapacitor devices does not have a significant deleterious impact on the optical emission properties of the few-layer MoS$_2$ crystals.

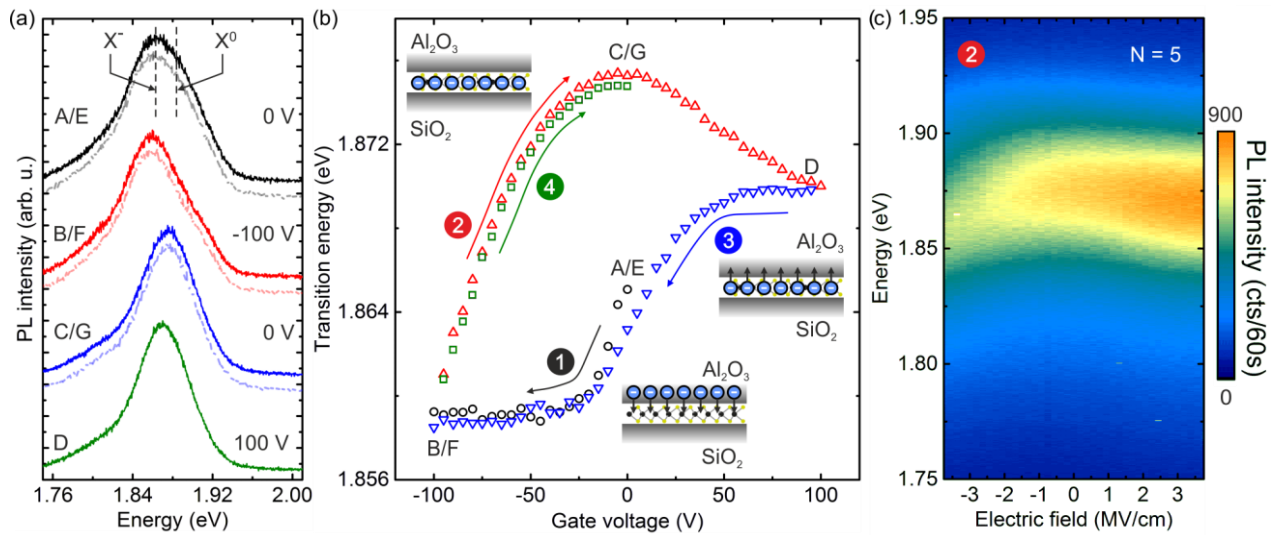

**Figure 2**. (a) Typical PL spectra at selected gate voltages taken from positions A-G from the gate voltage sweep in panel (b) for an N = 5 MoS$_2$ crystal. The dashed lines indicate the emission from neutral excitons X$^0$ and charged excitons X$^-$ in spectrum A and E, respectively. Spectra A and E show a significantly enhanced emission from neutral excitons X$^0$ on the high-energy side whilst other spectra reveal a dominant emission from charged excitons X$^-$. (b) Typical hysteresis measurement of the A-exciton transition energy as a function of the applied gate voltage for an N = 5 MoS$_2$ crystal. The gate voltage is tuned from $V_G = 0$ V (A) to $V_G = -100$ V (B), from $V_G = -100$ V (B) to $V_G = 100$ V (D), from $V_G = 100$ V (D) to $V_G = -100$ V (F) and from $V_G = -100$ V (F) back to $V_G = 0$ V (G). The inset illustrates the transfer of electrons during the gate sweep. A-B: Electrons from the proximal $Al_2O_3$ layer are transferred into the $MoS_2$ crystal. B-D: The electrons remain in the $MoS_2$ crystal. D-E: Electrons are transferred into the proximal $Al_2O_3$ layer. (c) Typical low temperature photoluminescence spectra (taken from trace 2) of the A-exciton transition as a function of the applied electric field. The dashed line indicates the maximum transition energy of the A-exciton of $E_{A,max} \sim 1.876$ eV.

We continue by turning our attention to the exploration of the voltage dependence of the emission from mono- to few-layer $MoS_2$ crystals in the capacitor structure. For the remainder of this letter we focus exclusively on the A-exciton peak. Figure 2(a) shows typical PL spectra of an N = 5 $MoS_2$ micro-crystallite as a function of the applied gate voltage (Spectra for additional voltages for N = 5 and N = 1 can be found in the Supporting Information section S4). Here, $V_G$ is looped twice between - 100 V $\leq V_G \leq$ 100 V starting from $V_G$ = 0 V (A) and stopping at $V_G$ = 0 V (G) tracing the paths from 1 → 4 as shown by the corresponding extracted voltage dependent A-exciton transition energies E(F) in figure 2(b). We clearly observe a strong voltage dependent hysteresis of the transition energy which we attribute to charging effects of the dielectrics around the $MoS_2$ crystal, most probably by charging and decharging of the proximal $Al_2O_3$ layer with photogenerated electrons.[25] Taking a closer look at spectrum A in figure 2(a), we observe a significant emission, both, from the high- and low-energy side of the A-exciton which we attribute to recombination of neutral excitons $X^0$ and negatively charged excitons $X^-$, respectively.[26] By tuning to $V_G$ = - 100 V (B) we observe strongly reduced emission from $X^0$, most probably induced by electrons from the proximal $Al_2O_3$ layer that transfer into the $MoS_2$ crystal due to the applied negative top-gate voltage. Despite the peak shift, the dominant emission from $X^-$ remains unaltered when tuning to $V_G$ = 100 V as can be seen in spectra C and D, respectively. Spectrum E constitutes the initial state at $V_G$ = 0 V and has a form that is identical to spectrum A, where a significant enhancement of $X^0$ is observed. This enhancement reflects a reduction of the electron density in the $MoS_2$ crystal since they transfer into the proximal $Al_2O_3$ layer. The observed response is found to repeat for subsequent sweeps of $V_G$ as can be seen for spectra F and G, respectively. This spectral field dependence is consistently observed for all measured samples. Therefore, when sweeping from negative $V_G$ = - 100 V (B) to positive $V_G$ = 100 V (D) (trace 2), we can primarily probe the $X^-$ electro-optical response as a function of the applied gate voltage and, thus, electric field. Moreover, the $X^-$ luminescence at A/E is observed to be red-shifted with respect to C/G by $\Delta E = 10$ meV, an observation that is interpreted as the difference between the effective applied electric field at those two positions. A simple analysis indicates that an electric field of F = (2.88 ± 0.64) MVcm$^{-1}$ (see Supporting Information section S5) is already present at A/E. This additional built-in electric field is directed antiparallel to the applied external field direction and, therefore, most likely originates from a charge accumulation in the proximal $Al_2O_3$ layer. Figure 2(c) shows a false colour image of the resulting PL along path 2 in figure 2(a) as a function of the applied electric field and emission energy. We clearly observe a significant red shift of the maximum of the $X^-$ emission of the A-exciton of $E_{A,max}$ ~ 1.876 eV (indicated by the dashed line) for the biased device in comparison to the unbiased case. The emission from the A-exciton for F = - 3.75 MVcm$^{-1}$ (F = 3.75 MVcm$^{-1}$) is distinctly shifted to lower energies by $\Delta E$ > - 16 meV ($\Delta E$ > - 5 meV) as compared to the emission without external field. Notably, the minor changes of the PL intensity $I_A$ at $E_{A,max}$ of about ± 0.25 $I_A(E_{A,max})$ and, thus, the oscillator strength for |F| = 3.75 MVcm$^{-1}$ is observed throughout all measured samples, indicating only a weak perturbation of the transition oscillator strength.

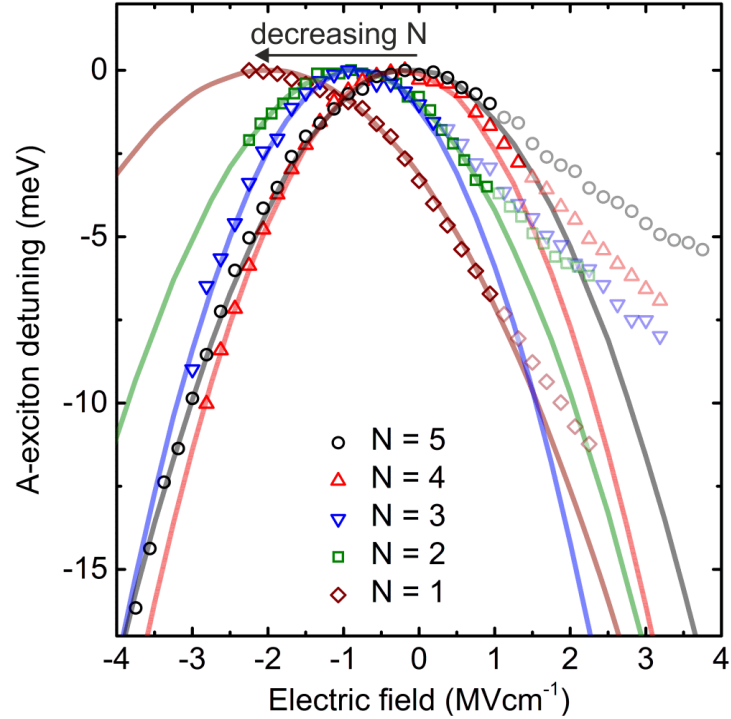

**Figure 3**. A-exciton detuning as a function of the applied electric field for N = 1, 2, 3, 4, 5 MoS$_2$ crystals and corresponding fits using the Stark formula (solid curves). The transparent data points at F > 0 MVcm$^{-1}$ were omitted for the fits as discussed in the main text.

We attribute the observed shift to the DC quantum confined Stark effect that has been widely used to tune the interband optical response of III-V quantum wells and dots.[27-30] In a simple model, we consider the exciton as a polarisable bound electron-hole-pair with finite spatial separation along the direction parallel to the basal plane of the crystal.[31,32] Polarisation of the exciton by the externally applied field combined with the possibility for a finite static exciton polarisation at F = 0, producing a nonzero excitonic dipole moment $p$ leads to an energy dependence of the exciton recombination energy of the form E(F) = E$_0$ − $p$ × F − $\beta$ × F$^2$ where E$_0$ is the exciton recombination energy at F = 0, $p$ the nonzero exciton dipole moment and $\beta$ the exciton polarisability. According to this simple expression, we would expect a quadratic dependence of the emission energy due to the exciton polarisability (quantum confined Stark effect) and a shift of the maximum transition energy E$_{A,max}$ away from F = 0 due to the nonzero dipole moment. To investigate the influence for a reducing number of layers N, we studied the electric field tuneable A-exciton emission for N = 1, 2, 3, 4, 5 regions of several flakes in different devices always following the path 2 in figure 2(b) to obtain reliable statistics. Typical results for the A-exciton field-dependence along path 2 for N = 1, 2, 3, 4, 5 regions of a MoS$_2$ crystal are presented in figure 3. We clearly observe the expected parabolic dependence of E(F) in figure 3 as indicated by the fits (solid curves) which we attribute to the typical signature of the quantum confined Stark effect. To obtain the best fit we omitted data points at F > 0 MVcm$^{-1}$ that evidently deviate from the parabolic detuning dependence due to the charging effects discussed in relation to figure 2(b). The data presented in figure 3 clearly exhibit a significant and systematic shift of the maximum transition energy $E_{A,max}$ towards more *negative* electric fields for a decreasing N, an observation to which we return below. The results presented in figure 3 strongly indicate that we probe the DC quantum confined Stark effect of X$^-$ in few-layer MoS$_2$ crystals facilitating the investigation of the exciton polarisability and nonzero excitonic dipole moment as a function of N.

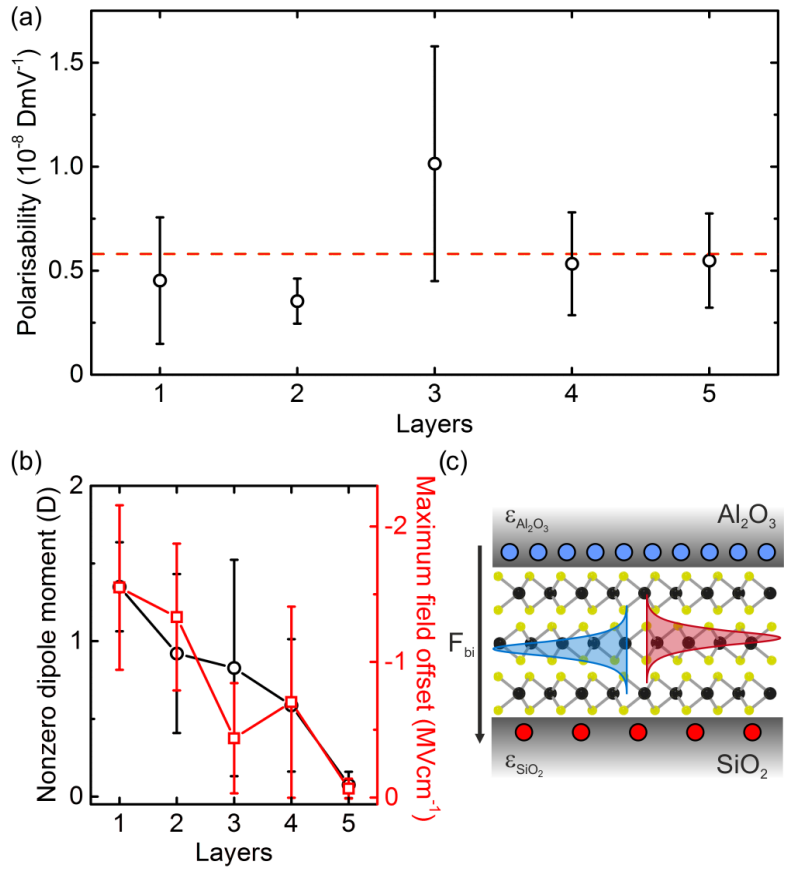

**Figure**. 4 (a) Mean values of the exciton polarisability $\bar{\beta}_N$ as a function of N (black circles) in units of Debye mV$^{-1}$. Also shown is the mean value of $\langle\bar{\beta}\rangle_N$ (dashed red line). (b) Mean values of the nonzero

exciton dipole moment as a function of N (left scale, black circles) and corresponding maximum field offset F$_{max}$ (right scale, red circles). (c) Schematic illustration of an N = 3 micro-capacitor device highlighting the difference of the surrounding polarisation charges (depicted as blue and red dots) due to the two different dielectrics $\varepsilon_{SiO_2}$ = 3.9 and $\varepsilon_{Al_2O_3}$ = 9.3 surrounding the TMDC. The resulting built-in field F$_{bi}$ is oriented antiparallel to the out-of-plane direction, directed from the Si-substrate to the top Au-contact. The electron and hole wave functions of optically generated excitons are strongly confined within individual layers and initially displaced.

We continue our discussion by deducing the exciton polarisability and nonzero exciton dipole moments from the measured field dependencies of many (> 7) samples. The results are presented in figure 4. The mean value of the A-exciton polarisability $\bar{\beta}_N$ is shown in figure 4(a) as a function of N. The exciton polarisability is scattered around a mean value of $\langle\bar{\beta}\rangle_N$ = (0.58 ± 0.25) × 10$^{-8}$ DmV$^{-1}$ (red dashed line) and is found to be independent of N within the experimental error. The polarisability in 2D systems, such as III-V quantum wells is well known to be strongly dependent on the width of the quantum well (w) according to $\beta \propto w^4$ since the wave function is delocalised over the width (w) of the QW. This would be expected to increase the polarisability by > one order of magnitude upon doubling the QW width[28] in complete contrast to our experiments observed. We attribute the observed constant polarisability to strong spatial confinement of the electron and hole wave functions within each individual monolayer in the N > 1 crystals. This interpretation is in full accord with ab-initio calculations for N = 1 to bulk MoS$_2$ crystals[33] that suggest a negligible wave function overlap of the A-exciton with adjacent atomic layers due to their large interlayer distance and weak van-der-Waals interaction. Furthermore, the experimentally obtained, extremely low values for the polarisability are reasonable by considering the very high exciton binding energies in TMDCs[32,34,35] which are on the order of a few hundred meV[33,36-38] for the A-exciton of MoS$_2$. Quantitatively, we can estimate the maximal effective relative displacement $r_{eff}^{max}$ of the electron and hole wave functions using $r_{eff}^{max}(F) = \frac{\langle\bar{\beta}\rangle_N}{e} \times \Delta F_{max}$ (with p = e × r ~ β × F), where ΔF$_{max}$ is the largest electric field applied with respect to the maximum of the transition energy. From the measured exciton polarisability and the largest $\Delta F_{max}$ that we could induce in our experiments we obtain an effective displacement of the electron and hole wave functions of $r_{eff}^{max} \sim$ (45 ± 20) pm for $\Delta F_{max}$ = 3.75 MVcm$^{-1}$. This value is more than one order of magnitude smaller than the thickness of a single atomic layer of MoS$_2$ $d_{(N = 1)} \sim$ 6.5 Å[8,39] and ~ one tenth of twice the covalent bond length between molybdenum and sulfur $d_{Mo-S} \sim$ 2.4 Å[39] in the crystal that represents an upper limit for $r_{eff}^{max}$. The observation of $r_{eff}^{max} \ll d_{(N = 1)}$, $d_{Mo-S}$ is expected considering the high exciton binding energies and large effective electron and hole masses.[33,36-38,40-42] Moreover, a simple analysis based on second order perturbation theory, as applied to quantum wells (see Supporting Information section S6) would suggest an effective width of $w < d_{(N = 1)}$ and the experimentally observed quantum confined Stark shift based on our extracted exciton polarisability. The small displacement at the maximum fields applied in our experiments further explains the weak reduction of PL intensity. The maximum observed energetic shifts of the transition energy $\Delta E \sim$ - 16 meV is small compared to the expected exciton binding energies and, hence, the applied electric field only weakly perturbs the oscillator strength of the transition in agreement with our observations in figure 2(c). We note that MoS$_2$ has been shown to exhibit a finite Stokes shift which is on the order of several tens of meV and, moreover, has been demonstrated to be doping dependent, resulting in a red-shift of the X$^-$ PL for an increased electron doping density.[26] As already depicted in figure 2(a) at position B/F we observe a weak residual contribution of X$^0$ luminescence which quickly vanishes when tuning to C/G which in contrast manifests in a blue-shift of the X$^-$ PL. Since we cannot fully rule out any contribution of a doping dependent Stokes shift which would result in a red-shift of the X$^-$ PL, the magnitude of the DC Stark effect observed in this work should be interpreted as reflecting a lower limit.

Finally, we discuss the nonzero dipole moment and its dependence on N. The extracted nonzero dipole moments obtained from fitting the Stark formula to E(F) are presented in figure 4(b) (black circles, left scale). The magnitude of the nonzero dipole significantly decreases with an increasing number of layers, with values of $p_{(N=1)}$ = (1.35 ± 0.29) D for N = 1 reducing to $p_{(N=5)}$ = (0.08 ± 0.08) D for N = 5. Similar as for the exciton polarisability, we would expect the nonzero excitonic dipole moment to be constant due to the strong exciton localisation in each individual atomic layer. Hence, we attribute the observed decreasing dipole moment and, hereby, the observed shift of the maximum transition energy $E_{A,max}$ to layer-dependent charging effects of the device and/or interaction between the outermost layers and the surrounding dielectrics that introduce a permanent excitonic dipole moment. By taking the derivative $\partial E/\partial F$ of the Stark formula and equating with zero we express $p$ and $\beta$ as a maximum field offset via $F_{max}$ = - $p$ ÷ (2 × $\beta$) which is shown in figure 4(b) (red circles, right scale). The maximum field offset decreases significantly for higher N similar to the nonzero dipole moment. The differing polarisation charge density in the neighbouring dielectrics lead to an initial displacement of the electron and hole wave function in an unbiased configuration as schematically illustrated in figure 4(c) for a N = 3 capacitor device. The differing dielectric constants $\varepsilon_{SiO_2} = 3.9$ and $\varepsilon_{Al_2O_3} = 9.3$ yield differing polarisation charge densities which locally act as a built-in-field $F_{bi}$ that is oriented antiparallel to the out-of-plane direction as given by the shift of the maximum transition energies and the external electric field. The reduction of the nonzero dipole moment observed for increasing N can thus be qualitatively understood as a compensation of the polarisation charges. A similar substrate screening effect was already observed for few-layer $MoS_2$ flakes prepared on $SiO_2$ by electrostatic force microscopy.[43] Prospective measurements in a symmetric dielectric environment or for a higher number (N > 5) of probed layers should compensate this screening effect in order to be able to give clear indications regarding the intrinsic A-exciton dipole moment.

In summary, we realised a micro-capacitor device which allowed the application of high electric fields perpendicular to the basal plane of few-layered $MoS_2$ crystals while simultaneously facilitating optical access to the material. Studying the A-exciton emission of N = 1 to N = 5 $MoS_2$ crystals as a function of the applied electric field, we observe clear evidence of a DC quantum confined Stark effect with a significant tuneability of $\Delta E > -16$ meV. This allows us to deduce a very low layer independent polarisability $\langle \bar{\beta} \rangle_N = (0.58 \pm 0.25) \times 10^{-8}$ DmV$^{-1}$ consistent with the strongly confined nature of the A-exciton where the electron and hole wave functions are spatially localised within each individual layer. The weakly perturbed A-exciton oscillator strength for the high electric fields applied in our experiments further renders MoS$_2$ as an excellent candidate for utilising the DC Stark effect for optoelectronic devices. An external built-in-field originating from the differing polarisation charge densities of the neighbouring dielectrics and/or interaction of the outermost layers with the neighbouring dielectrics leads to an increasing offset of the Stark parabola for thinner crystals which prevents a clear interpretation on the intrinsic exciton dipole moment. Our work further pushes the focus of atomically thin two-dimensional materials towards novel devices enabling to exploit the voltage-controlled intriguing photo-physical properties of few-layered TMDCs. Furthermore, the demonstration that electronic structure can be tuned using a DC electric field could have important implication for the fraction of luminescence recombining at the direct and indirect band gap.[44]

ASSOCIATED CONTENT

**Supporting Information**. Layer-dependent Raman measurements for validating the MoS$_2$ layer number, device capacitance calculation, typical photocapacitor transfer curve, finite-element simulation of the applied electric field at the MoS$_2$ micro-crystallite, gate voltage dependent PL of N = 1 and N = 5 MoS$_2$, electric field approximation, second order perturbation model for quantum wells.

AUTHOR INFORMATION

**Corresponding Author**

[†]Correspondence to kaniber@wsi.tum.de or finley@wsi.tum.de

**Author Contributions**

J.W., J.K. and J.J.F. conceived and designed the experiments, J.K., J.W. and F.H. prepared the samples, J.K. and J.W. performed the optical measurements, J.K. and J.B. performed the electrical measurements, A.R., J.K. and J.W performed finite-element simulations, J.K. analysed the data, J.K., J.W., M.K., K.M. and J.J.F. wrote the paper. All authors reviewed the manuscript. *These authors contributed equally.

ACKNOWLEDGMENT

Supported by Deutsche Forschungsgemeinschaft (DFG) through the TUM International Graduate School of Science and Engineering (IGSSE). We gratefully acknowledge financial support of the German Excellence Initiative via the Nanosystems Initiative Munich and the PhD programme ExQM of the Elite Network of Bavaria. A.R. acknowledges support from the Technische Universität München – Institute for Advanced Study, funded by the German Excellence Initiative. K.M. acknowledges support from the Alexander von Humboldt Foundation.

ABBREVIATIONS

TMDC, transition metal dichalcogenide; PL, photoluminescence;